\ifx\documentclass\undefined  
  \documentstyle[11pt,a4]{article}
\else
  \documentclass[12pt,a4paper]{article}
\fi

\begin{document}
\begin{center}
{\bf Neutron Scattering from Magnetic Excitations in $\bf Bi_2 Sr_2 Ca
Cu_2 O_{8+\delta}$}

\vspace{.5in}

\noindent  H.F. Fong$^{(1)}$, P. Bourges$^{(2)}$, Y. Sidis$^{(2)}$,
L.P. Regnault$^{(3)}$, A. Ivanov$^{(4)}$,\\
G.D. Gu$^{(5)}$, N. Koshizuka$^{(6)}$, and B. Keimer$^{(1,7)}$

\end{center}

\vspace{.5in}

\begin{tabular}{lp{4.5in}}
1& Department of Physics, Princeton University, Princeton, NJ 08544,
USA\\
2& Laboratoire L\'eon Brillouin, CEA-CNRS, CE Saclay, 91191 Gif sur 
Yvette, France\\
3& CEA Grenoble, D\'epartement de Recherche Fondamentale sur 
la mati\`ere Condens\'ee, 38054 Grenoble cedex 9, France\\
4 & Institut Laue-Langevin, 156X, 
38042 Grenoble Cedex 9, France\\
5 & Department of Advanced Electronic Materials, School of Physics,
University of New South Wales, Sydney 2052, Australia\\
6 & SRL/ISTEC, 10-13, Shinonome 1-chome, Koto-ku, Tokyo 135, Japan\\
7 & Max-Planck-Institut f\"ur Festk\"orperforschung, 70569 Stuttgart,
Germany
\end{tabular}
                                             
\clearpage

\noindent {\bf Many physical properties of the copper oxide high temperature
superconductors appear to defy the conventional (one-electron) theory of
metals, and the development of new theories incorporating strong electron 
correlations
is currently at the forefront of condensed matter physics. 
Inelastic neutron scattering provides incisive information about
collective magnetic excitations that is required to guide this effort.
Such measurements have thus far proven possible for only two of the many
families of high temperature superconductors, $\bf La_{2-x} Sr_x CuO_4$
\cite{kastner} and $\bf YBa_2 Cu_3 O_{6+x}$ \cite{tranquada}-\cite{bourges97_1},
because suitably large single crystals of other copper oxide compounds could not 
be grown. While the magnetic spectra of both
materials bear certain similarities, there are also pronounced differences
that have hampered a unified description of the spin dynamics in the
copper oxides. In particular, a sharp resonant spin excitation dominates the
spectrum in the superconducting state of $\bf YBa_2 Cu_3 O_{6+x}$ 
\cite{rossat}-\cite{bourges97_1}, but
is not found in $\bf La_{2-x} Sr_x CuO_4$ \cite{kastner}. Here we report
the discovery of a magnetic
resonance peak in the superconducting state of a different copper oxide system, 
$\bf Bi_2 Sr_2 Ca Cu_2 O_{8+\delta}$, made possible by the synthesis of
a sizable single crystal and modern neutron optics. This provides
evidence of the generality of this unusual phenomenon among the copper oxides 
and greatly
extends the empirical basis for its theoretical description.}

\vspace{.2in}

The magnetic resonance peak observed in optimally doped $\rm YBa_2 Cu_3 O_{6+x}$
(superconducting transition
temperature $\rm T_c = 93K$) is a sharp collective mode that occurs at an
energy of 40 meV and two-dimensional wave vector ${\bf Q}_{||}=
(\pi/a,\pi/a)$, where $a$ is the nearest-neighbor Cu-Cu distance
\cite{rossat}-\cite{fong96}. Its intensity decreases
continuously with increasing temperature and vanishes above $\rm T_c$.
In underdoped $\rm YBa_2 Cu_3 O_{6+x}$, the mode energy decreases
monotonically with decreasing hole concentration 
\cite{fong97}-\cite{bourges97_1}, while the
superconducting energy gap is believed to remain approximately constant. 
Such a collective mode has not been observed in conventional superconductors. 
Several
microscopic mechanisms have been proposed, ranging from band structure
singularities \cite{bulut}-\cite{abrikosov} to antiferromagnetic 
spin fluctuations \cite{liu}-\cite{morr} and
interlayer pair tunneling \cite{chakravarty}. In all of these models, 
the interactions that give rise to
the resonant mode are the same that cause pairing of the electrons in the
superconducting state, so that this phenomenon provides a direct clue to the
mechanism of high temperature superconductivity.
According to another recent theory, the resonance mode is a
manifestation of a new symmetry group linking antiferromagnetism and
superconductivity \cite{zhang,demler}. In the framework of this model,
the spectral weight of the resonance peak is in fact directly proportional to 
the
superconducting condensation energy \cite{scalapino,zhang1}.

A very sensitive test of these
disparate models is whether
they are capable of providing detailed descriptions of both the inelastic
neutron scattering (INS) results and those of angle-resolved
photoemission spectroscopy (ARPES), a complementary 
momentum-resolved experimental technique that primarily probes 
single-electron excitations. 
By far the best ARPES data have been obtained on 
$\rm Bi_2 Sr_2 Ca Cu_2 O_{8+\delta}$ \cite{ding,shen}, a material for
which no INS data
have been available. This situation, which has precluded a direct,
quantitative comparison of the results of both techniques, is remedied by
the present study.

A look at the chemical bonding in the copper oxides reveals the origin of these
experimental difficulties. In $\rm Bi_2 Sr_2 Ca Cu_2 O_{8+\delta}$, for
instance, the $\rm CuO_2$ planes are separated by several charge
reservoir layers that interact only weakly in the
$\hat{c}$-direction (perpendicular to the layers). $\rm Bi_2 Sr_2 Ca Cu_2
O_{8+\delta}$ therefore cleaves easily
along the $\rm CuO_2$ layers, which greatly facilitates surface-sensitive
techniques such as ARPES. This same property is responsible for the lack
of INS data on this compound. In order to take advantage of the momentum
resolution of INS, the signal from a large volume of material has to be
coherently superposed, which is only possible if a single crystal is
available. However, because of the weak bonding along $\hat{c}$, 
$\rm Bi_2 Sr_2 Ca Cu_2 O_{8+\delta}$ crystals typically grow as thin
plates with volumes much too small for INS. The present study was
performed on a comparatively large single crystal of volume $\rm 10
\times 5 \times 1.2 \; mm^3$ and mosaicity
$\sim 1^\circ$, synthesized by the floating zone technique as 
described previously \cite{gu}. Its
superconducting transition temperature (measured by SQUID
magnetometry) is $\rm T_c$=91K, with a width of 2K.

The crystal was mounted
in an aluminum can filled with helium exchange gas and attached to the
cold finger of a closed cycle helium refrigerator.
The INS experiments were performed at the spectrometers 2T at the 
Laboratoire L\'eon Brillouin, Saclay, France, and IN8 at the
Institut Laue-Langevin in Grenoble, France. Both spectrometers use
neutron optics that focus the beam both parallel and perpendicular to the
scattering plane, with a resulting gain in neutron flux on the sample that
proved crucial for this experiment. Identical configurations [Cu(111)
monochromator, PG(002) analyser, 35 meV fixed final energy] were used
at both spectrometers, so that the data can be directly compared after
correcting for the different neutron fluxes at both reactors. The
inelastic cross section was put on an absolute scale by a calibration
against the incoherent scattering of vanadium. The comparison with
$\rm YBa_2 Cu_3 O_7$ (below) is also facilitated by the availability of data
taken on the same spectrometer under the same experimental conditions
\cite{bourges96}.

As our $\rm Bi_2 Sr_2 Ca Cu_2 O_{8+\delta}$ crystal has a similar
superconducting transition temperature as optimally doped 
$\rm YBa_2 Cu_3 O_{6+x}$, we began our search for the resonance mode in
an energy range around 40 meV. Fig. 1a shows scans of the neutron
intensity as a function of the in-plane wave vector ${\bf Q}_{||}$ at a
constant excitation energy of 43 meV. A featureless 
background, mostly due to multiphonon processes commonly observed 
in materials with large unit cells, is
present at all temperatures. While no signal can be identified above this
flat background at T=100K, a pronounced peak centered at ${\bf
Q}_{||}=(\pi/a,\pi/a)$ appears at lower temperature. The peak amplitude
measured at symmetry-equivalent points in reciprocal space decreases
with increasing wave vector, providing reassurance that the
peak is of magnetic origin, and it compares well to the magnetic form factor
of copper measured in antiferromagnetically long-range ordered copper oxides.

Indeed, this behavior is one of the hallmarks of
the magnetic resonance peak in the superconducting state of $\rm YBa_2
Cu_3 O_{6+x}$. Another characteristic feature of the resonance peak is
revealed by a scan of the wave vector ${\bf Q}_{\perp}$ perpendicular to
the $\rm CuO_2$ planes. As shown in Fig. 1b, the additional low
temperature intensity is sinusoidally modulated as a function of ${\bf
Q}_{\perp}$, with a periodicity equal to the inverse of the distance of two
adjacent $\rm CuO_2$ planes. (Because of kinematical restrictions at low
$Q$ and phonon contamination at high $Q$, only one of the maxima of the
sinusoid is accessible.) The modulation is a consequence of strong
magnetic interactions between adjacent layers, whose microscopic nature
has been the focus of much debate in the context of the magnetic
resonance peak \cite{mazin}-\cite{zhang}. 

We now proceed to demonstrate the presence of the remaining signatures
of the magnetic resonance peak in the $\rm Bi_2 Sr_2 Ca Cu_2
O_{8+\delta}$ data. Fig. 2 shows a scan of the additional low temperature
intensity as a function of the excitation energy, with the wave vector
fixed at the maximum in both ${\bf Q}_{||}$ and ${\bf Q}_{\perp}$. The
intensity is peaked around an energy of $\sim 43$ meV, with a width of
$\sim$ 10-15 meV. Finally, Fig. 3 shows the temperature dependence of
the peak amplitude which vanishes above the superconducting transition
temperature, 
to within the experimental uncertainty.

The magnetic excitation spectra of $\rm Bi_2 Sr_2 Ca Cu_2
O_{8+\delta}$ and $\rm YBa_ 2 Cu_3 O_7$ thus exhibit an unmistakable
similarity. In the superconducting state, 
the magnetic intensity is sharply concentrated around a single point in
energy and wave vector; in the normal state, the intensity is either too
broad or too weak to be observable above background. There is also no
indication of magnetic intensity above background level at other energies
or wave vectors. In particular, an extensive search for magnetic
excitations at 10 meV, an energy below the onset of optical phonon scattering 
where 
magnetic excitations, if present, should be relatively easy to observe,
has thus far been fruitless in $\rm Bi_2 Sr_2 Ca Cu_2 O_{8+\delta}$. 
In both $\rm YBa_ 2 Cu_3 O_7$ and
$\rm Bi_2 Sr_2 Ca Cu_2 O_{8+\delta}$, the resonance peak is therefore by
far the most prominent feature in the magnetic excitation spectrum.

A further important comparison
between both compounds is made possible by a calibration of the absolute
neutron cross section against a vanadium standard. Expressed in the same
units as in Ref. \cite{bourges97_2}, the energy-integrated spectral weight
of the resonance peak at ${\bf Q}_{||}=(\pi/a,\pi/a)$ is $\int d (\hbar
\omega) \chi'' ({\bf Q},\omega) = (1.9 \pm 1) \mu_B^2$, where $\chi''$ is 
the imaginary part of the dynamical spin
susceptibility. (The large error bar is mostly due to the uncertainty in the
energy width, Fig. 2). 
To within the experimental error, this is identical to $\rm 1.6 \mu_B^2$
determined
in $\rm YBa_2 Cu_3 O_7$ \cite{fong96,zn}. Surprisingly, the width of
the resonance in ${\bf Q}_{||}$ is much larger in $\rm Bi_2
Sr_2 Ca Cu_2 O_{8+\delta}$ ($0.52 \rm{\AA}^{-1}$ full width at half
maximum) than in $\rm YBa_2 Cu_3 O_7$ ($0.25 \rm{\AA}^{-1}$). If averaged over 
the 
two-dimensional Brillouin zone, in addition to integrating
over $\hbar \omega$, the resonance spectral weight is therefore
clearly larger in $\rm Bi_2 Sr_2 Ca Cu_2 O_{8+\delta}$ ($\int d {\bf Q}_{||} d 
(\hbar
\omega) \chi'' ({\bf Q},\omega) / \int d {\bf Q}_{||}  = 0.23 \mu_B^2$) 
than in $\rm YBa_2 Cu_3 O_7$ ($0.043 \mu_B^2$ \cite{zn}).

Such quantitative comparisons between different materials are
indispensable for a microscopic, quantitative description of the origin 
of the magnetic resonance peak. In the framework of the models
proposed for the resonance peak, it should now be possible to relate the
different {\bf Q}-widths measured in $\rm YBa_2 Cu_3 O_7$ and $\rm Bi_2 Sr_2 Ca
Cu_2 O_{8+\delta}$
to their respective Fermi surfaces as measured by ARPES. (The difference in the 
energy widths
may be more difficult to interpret, as recent experiments on Zn-substituted
$\rm YBa_2 Cu_3 O_7$ indicate that it is very sensitive to disorder \cite{zn}. 
Incidentally,
this sensitivity may account for the absence of the resonance peak in $\rm 
La_{2-x} Sr_x CuO_4$,
where disorder due to Sr donors is significant. The {\bf Q}-width,
on the other hand, is not strongly affected by impurities.) There have also
been attempts to relate more subtle features of the ARPES spectra, such as the 
well-known
``dip'' in the spectral function, to the collective spin dynamics  
\cite{norman,schrieffer}. These features are only clearly apparent in the
ARPES work on
$\rm Bi_2 Sr_2 Ca Cu_2 O_{8+\delta}$, and our INS data will now allow
such arguments
to be refined and put on a quantitative footing. The same applies to recent
models that propose a direct relation between the spectral weight of the
resonance peak and 
the superconducting condensation energy \cite{scalapino,zhang1}.

Our study opens 
the way for a variety of further neutron experiments, in particular in the
overdoped
regime that is easily accessible in $\rm Bi_2 Sr_2 Ca Cu_2 O_{8+\delta}$
over a wide range 
of hole concentrations. It also leaves open questions that can only be
answered by neutron 
scattering work on other families of high temperature superconductors.
For instance, as both $\rm YBa_2 Cu_3 O_7$ and $\rm Bi_2 Sr_2 Ca Cu_2
O_{8+\delta}$ 
are bilayer materials, 
the present study does not provide further insight into the role of
interlayer interactions 
in the resonance peak. Most importantly, observation of the resonance peak
in $\rm Bi_2 Sr_2 Ca Cu_2 O_{8+\delta}$ rules out the possibility that this
phenomenon is due to a conspiracy of structural or chemical parameters peculiar 
to
$\rm YBa_2 Cu_3 O_7$. Rather, it is an intrinsic feature of the copper oxides 
whose explanation
must be an integral part of any theory of high temperature
superconductivity. 

\vspace{.1in}

{\bf Acknowledgments}. We acknowledge useful discussions
with A.H. Moudden. The work at Princeton University was supported by the
MRSEC program of the National Science Foundation, and by the Packard and
Sloan Foundations.

\clearpage
\subsection*{Figure Captions}
\begin{enumerate}
\item Constant-energy scans, at excitation energy 43 meV, of the neutron
scattering intensity as a
function of the wave vectors (a) ${\bf Q}_{||}$ and (b) 
${\bf Q}_{\perp}$ parallel and perpendicular to the $\rm CuO_2$ planes,
respectively.
The wave vector $\rm {\bf Q}=(H,K,L)$ is given in reciprocal lattice units
(r.l.u.), that is, in units
of the reciprocal lattice vectors $a^* \sim b^* = 1.64 {\rm \AA}^{-1}$ and
$c^* = 0.20
{\rm \AA}^{-1}$. In these units, ${\bf Q}_{||} = (0.5,0.5)$ corresponds to the
antiferromagnetic wave vector $(\pi/a,\pi/a)$. The points in (a) are raw
data 
taken at temperatures 10K ($\rm < T_c$)
and 100K ($\rm > T_c$). In (b), the intensity at 100K was subtracted from
that at 10K. 
The line in (a) is the result of a fit to a Gaussian profile while that in (b)
is the function 
$\sin ^2 (\pi z_{\rm Cu} L)$, where $z_{\rm Cu} = 0.109$ is the distance
between nearest-neighbor
copper atoms in the $\hat{c}$-direction perperdicular to the planes,
expressed as a fraction of the
lattice parameter $c = 30.8 {\rm \AA}$. This modulation is  
due to magnetic interactions between adjacent $\rm CuO_2$ layers 
and has the same periodicity and phase as in $\rm YBa_2 Cu_3 O_7$. The
bars represent the experimental wave vector resolution (full width at half
maximum).

\item Difference spectrum of the neutron intensities at T=10K ($\rm <
T_c$) 
and T=100K ($\rm > T_c$), at wave vector ${\bf Q} = (0.5,0.5,-14)$.
The bar respresents the instrumental energy resolution, the line is a
guide-to-the-eye.

\item 
Temperature dependence of the neutron intensity at energy 43 meV and
wave vector 
${\bf Q} = (0.5,0.5,-14)$. The intensity falls to background level above
$\rm T_c = 91K$
(Fig. 1). The line is a guide-to-the-eye.
\end{enumerate}

\clearpage

\begin{thebibliography}{99}

\bibitem{kastner} Kastner, M.A,, Birgeneau, R.J., Shirane, G. \& Endoh, Y.
Magnetic, transport, and optical properties of monolayer copper oxides,
{\it Rev. Mod. Phys.} {\bf 70}, 897-928 (1998). 

\bibitem{tranquada} Tranquada, J.M., Shirane, G., Keimer, B., Shamoto, S. \&
Sato, M.
Neutron Scattering Study of Magnetic Excitations in $\rm YBa_2 Cu_3
O_{6+x}$,
{\it Phys. Rev.} B{\bf 40}, 4503-4516 (1989).

\bibitem{rossat} Rossat-Mignod, J. {\it et al.} 
Neutron Scattering Study of the $\rm YBa_2 Cu_3 O_{6+x}$ System,
{\it Physica} C{\bf 185-189}, 86-92 (1991).
 
\bibitem{mook93} Mook, H.A, Yethiraj, M., Aeppli, G., Mason, T.E. \&
Armstrong, T. Polarized Neutron Determination of the Magnetic Excitations
in
$\rm YBa_2 Cu_3 O_7$, {\it Phys. Rev. Lett.} {\bf 70}, 3490-3493 (1993).

\bibitem{fong95} Fong, H.F. {\it et al.} Phonon and Magnetic Neutron
Scattering at 41 meV in
$\rm YBa_2 Cu_3 O_7$, {\it Phys. Rev. Lett.} {\bf 75}, 316-319 (1995).

\bibitem{bourges96} Bourges, P., Regnault, L.P., Sidis, Y., \& Vettier, C.
Inelastic-neutron-scattering study of antiferromagnetic fluctuations in
$\rm YBa_2Cu_3O_{6.97}$, {\it Phys. Rev.} B{\bf 53}, 876-885 (1996).

\bibitem{fong96} Fong, H.F., Keimer, B., Reznik, D., Dogan, F. \&
Aksay, I.A. Polarized and unpolarized neutron-scattering study of the
dynamical
spin susceptibility of $\rm YBa_2Cu_3O_7$, {\it Phys. Rev.} B{\bf 54},
6708-6720 (1996).

\bibitem{fong97} Fong, H.F., Keimer, B., Milius, D.L. \& Aksay, I.A.
Superconductivity-induced anomalies in the spin excitation spectra of
underdoped 
$\rm YBa_2Cu_3O_{6+x}$, {\it Phys. Rev. Lett.} {\bf 78}, 713-716 (1997).

\bibitem{dai} Dai, P., Yethiraj, M., Mook, H.A., Lindemer, T.B. \& Dogan, F.
Magnetic dynamics in underdoped $\rm YBa_2Cu_3O_{7-x}$: Direct
observation of a superconducting gap, {\it Phys. Rev. Lett.} {\bf 77}, 5425-
5428 (1996).

\bibitem{bourges97_1} Bourges, P., {\it et al.} Shifting of the magnetic-
resonance 
peak to lower energy in the superconducting state of underdoped $\rm
YBa_2Cu_3O_{6.8}$,
{\it Europhysics Lett.} {\bf 38}, 313-318 (1997).

\bibitem{bulut} Bulut, N. \& Scalapino, D.J. Neutron scattering from a
collective spin 
fluctuation mode in a $\it CuO_2$ bilayer, {\it Phys. Rev.} B{\bf 53},
5149-5152 (1996).

\bibitem{blumberg} Blumberg, G., Stojkovic, B.P. \& Klein, M.V.
Antiferromagnetic 
excitations and van Hove singularities in $\rm YBa_2Cu_3O_{6+x}$, {\it
Phys. Rev.} 
B{\bf 52}, 15741-15744 (1995).

\bibitem {abrikosov} Abrikosov, A.A. Neutron peak in the extended-saddle-
point model 
of high-temperature superconductors, {\it Phys. Rev.} B{\bf 57}, 8656-
8661 (1998).

\bibitem{liu} Liu, D.Z., Zha, Y. \& Levin, K. Theory of Neutron Scattering in
the Normal and 
Superconducting States of $\rm YBa_2 Cu_3 O_{6+x}$, {\it Phys. Rev.
Lett.}
{\bf 75}, 4130-4133 (1995).

\bibitem{mazin} Mazin, I.I. \& Yakovenko, V.M. Neutron scattering and
superconducting 
order parameter in $\rm YBa_2Cu_3O_7$, {\it Phys. Rev. Lett.}
{\bf 75}, 4134-4137 (1995).

\bibitem{onufrieva1} Onufrieva, F. \& Rossat-Mignod, J. Theory of spin
dynamics in high-$\rm T_c$ copper oxide superconductors: Application to
neutron scattering, {\it Phys. Rev.} B{\bf 52}, 7572-7603 (1995).

\bibitem{onufrieva2} Onufrieva, F. Evidence for $d_{x^2-y^2}$ Symmetry of
the Superconducting
Order Parameter in YBCO from Neutron-Scattering Data, {\it Physica}
C{\bf 251}, 348-354 (1995).

\bibitem{millis} Millis, A.J. \& Monien, H. Bilayer coupling in the yttrium-
barium 
family of high-temperature superconductors, {\it Phys. Rev.} B{\bf 54},
16172-16178 (1996). 

\bibitem{morr} Morr, D.K. \& Pines, D. The resonance peak in cuprate
superconductors,
{\it Phys. Rev. Lett.} {\bf 81}, 1086-1089 (1998).

\bibitem{chakravarty} Yin, L., Chakravarty, S. \& Anderson, P.W.
The neutron peak in the interlayer tunneling model of high
temperature superconductors, {\it Phys. Rev. Lett.} {\bf 78}, 3559-3562
(1997).

\bibitem{zhang} Zhang, S.C. A unified theory based on SO(5) symmetry of
superconductivity and
antiferromagnetism, {\it Science} {\bf 275}, 1089-1096 (1997).

\bibitem{demler} Demler, E., Kohno, H. \& Zhang, S.C. $\pi$-excitation of
the t-J model,
{\it Phys. Rev.} B{\bf 58}, 5719-5730 (1998).

\bibitem{scalapino} Scalapino, D.J. \& White, S.R. The Superconducting
Condensation Energy and an Antiferromagnetic Exchange Based Pairing
Mechanism, preprint cond-mat/9805075 at http://xxx.lanl.gov.

\bibitem{zhang1} Demler, E. \& Zhang, S.C. Quantitative Determination of a
Microscopic Mechanism of High Tc Superconductivity, 
{\it Nature} {\bf 396}, 733-735 (1998).

\bibitem{ding} Norman, M.R. {\it et al.} Destruction of the Fermi surface
underdoped 
high-$\rm T_c$ superconductors, {\it Nature} {\bf 392}, 157-160 (1998).

\bibitem{shen} Shen, Z.X. {\it et al.} Temperature-induced momentum-dependent 
spectral weight 
transfer in $\rm Bi_2Sr_2CaCu_2O_{8+\delta}$, {\it Science} {\bf 280},
259-262 (1998).

\bibitem{gu} Gu, G.D., Takamuku, K., Koshizuka, N. \& Tanaka, S.
Large single crystal Bi-2212 along c-axis prepared by floating zone
method, {\it J. Crystal Growth} {\bf 130}, 325-329 (1998).

\bibitem{bourges97_2} Bourges, P. {\it et al.} High-energy spin
excitations in $\rm YBa_2Cu_3O_{6.5}$, {\it Phys. Rev.} B{\bf 56},
R11439-R11442 (1997). 

\bibitem{zn} Fong, H.F. {\it et al.} Effect of nonmagnetic impurities on
the magnetic resonance peak in $\rm YBa_2 Cu_3 O_7$, 
{\it Phys. Rev. Lett.} (in press); also preprint
cond-mat/9812047 at http://xxx.lanl.gov.

\bibitem{norman} Norman, M.R. \& Ding, H. Collective modes and the
superconducting-state spectral function of $\rm Bi_2Sr_2CaCu_2O_8$,
{\it Phys. Rev.} B{\bf 57} R11089-R11092 (1998).

\bibitem{schrieffer} Shen, Z.X. \& Schrieffer, J.R. Momentum, temperature,
and doping dependence of photoemission lineshape and implications for the
nature of the pairing potential in high-$\rm T_c$ superconducting
materials, {\it Phys. Rev. Lett.} {\bf 78}, 1771-1774 (1997). 

\end{thebibliography}
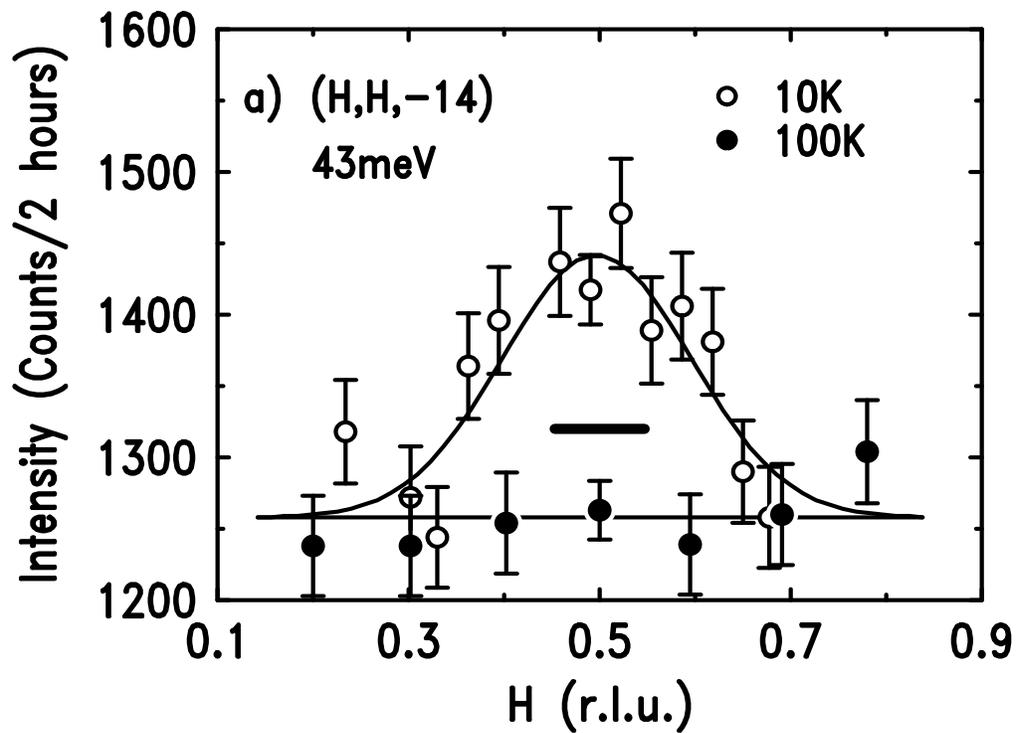
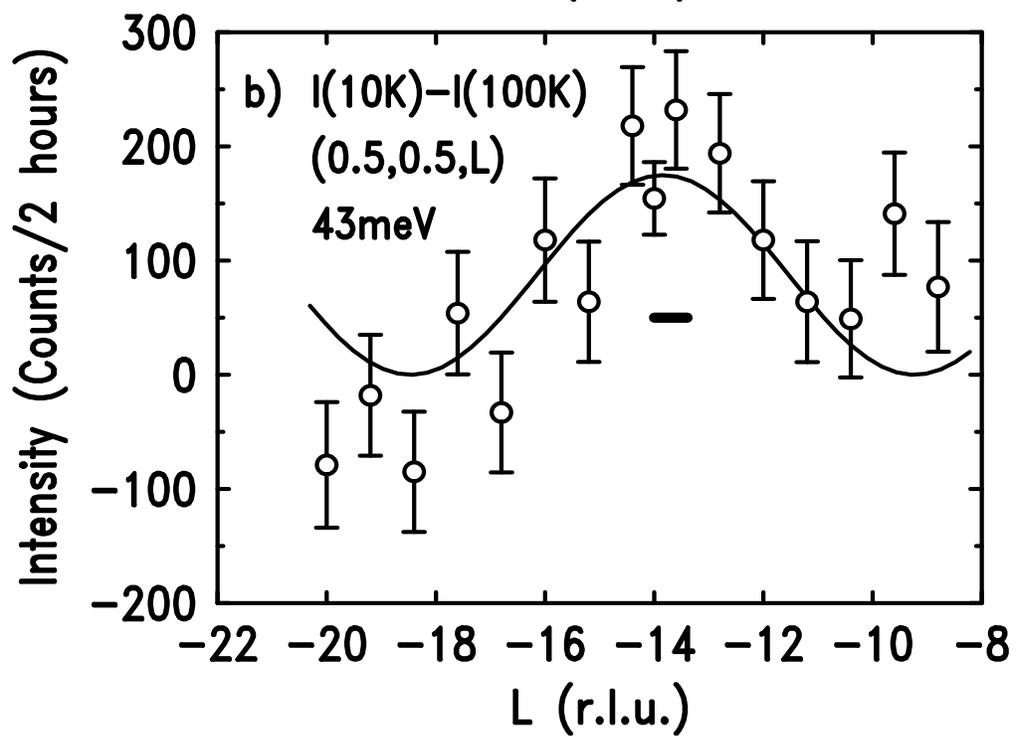
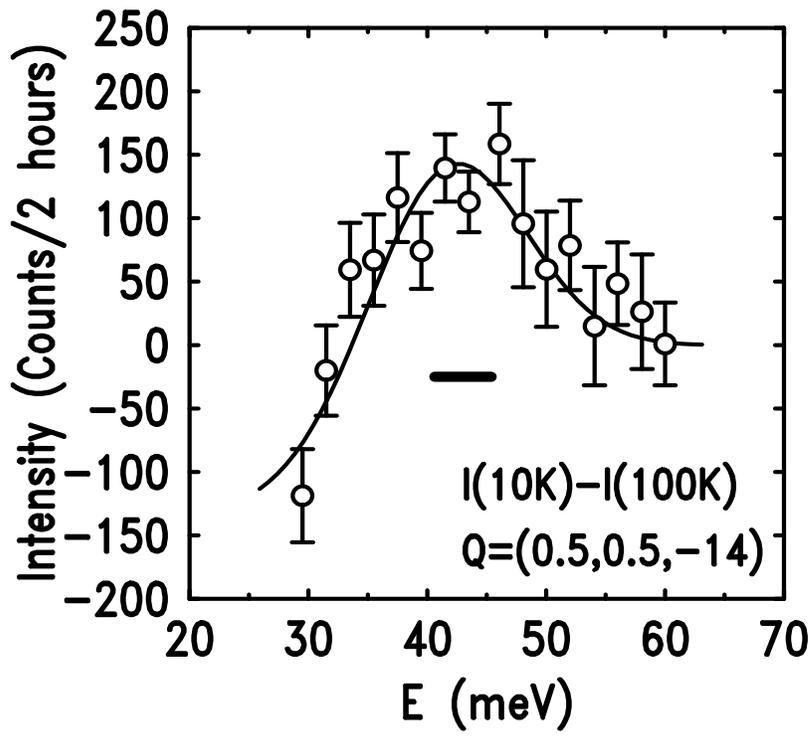
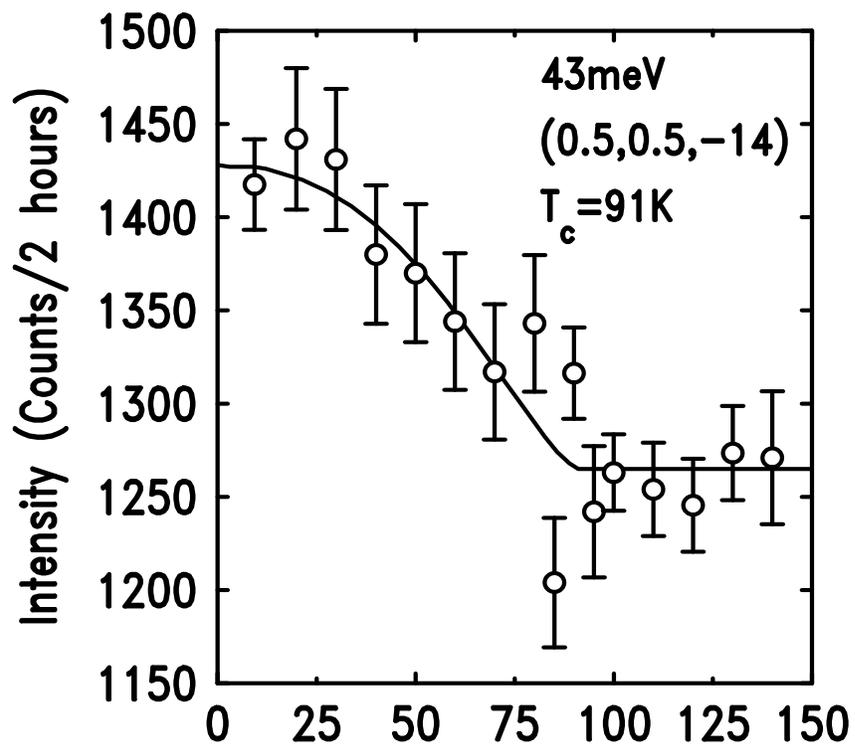
\end{document}